\documentclass[12pt]{iopart}

\usepackage{epsf}

\begin{document}

\title{A Supersymmetry approach to billiards with randomly distributed scatterers}

\author{H.-J. St\"ockmann}

\address{Fachbereich Physik der Philipps-Universit\"at Marburg,
D-35032 Marburg, Germany\\[2ex]
E-mail: stoeckmann@physik.uni-marburg.de}

\begin{abstract}
The density of states for a chaotic billiard with randomly distributed point-like
scatterers is calculated, doubly averaged over the positions of the impurities and the
shape of the billiard. Truncating the billiard Hamiltonian to a $N\times N$ matrix, an
explicit analytic expression is obtained for the case of broken time-reversal symmetry,
depending on rank $N$ of the matrix, number $L$ of scatterers, and strength of the
scattering potential. In the strong coupling limit a discontinuous change is observed in
the density of states as soon as $L$ exceeds $N$.
\end{abstract}


\newpage

\renewcommand{\Re} {\mathrm{Re}}

\renewcommand{\Im} {\mathrm{Im}}

\newcommand{\eb} {{\bf1}}

\newcommand{\Eb} {{\bf E}}

\newcommand{\Hob} {{\bf H_0}}

\newcommand{\Hb} {{\bf H}}

\newcommand {\xf} {{\bf x}}

\eqnobysec

\section{Motivation}

Experiments with classical waves have become a very versatile tool to study localization
due to disorder. In particular the experiments by Lagendijk and coworkers (Wiersma \etal
1997) on the localization of light in powders, and by the Genack group on the
localization of microwaves in disordered metallic spheres (Chabanov and Genack 2001) have
to be mentioned (for a review of these types of experiments see Soukoulis 1996). Moreover
microwave techniques are able to study spatially resolved field distributions in
disordered systems of linear dimensions in the order of some 10\,cm (Kudrolli \etal 1995,
St\"ockmann \etal 2001). Such quantities are inaccessible in electronic quantum dot
systems of submicron size (except for the recent experiments by Topinka \etal (2001)).
With increasing frequency one typically observes a transition from localized to
delocalized wave functions, depending on the number of scatterers and the strength of the
scattering potential. Pulse propagation can be studied as well by microwave techniques as
has been shown by Stein \etal (1995). All quantities of interest are thus experimentally
accessible in disordered systems, including conductivity, localization-delocalization
transitions, pulse propagation, transition from the ballistic to the diffusive regime,
and so on.

On the theoretical side the situation is less favourable. Though there is a
vast amount of literature on disordered systems already in the seventies and
the eighties of the last century (see e.\,g. Anderson 1978, Lee and
Ramakrishnan 1985 for reviews), there is as yet no theory available covering
the complete range from the localized to the delocalized regime. Today the
standard approach to study disordered systems uses supersymmetry techniques
to arrive at Efetov's non-linear $\sigma$ model (Efetov 1983). It has the
serious draw-back that the occurring supersymmetric variables are field
variables depending on the position. Only in the zero mode approximation,
where the position dependencies are neglected, the model can be solved
exactly and reproduces random matrix theory. This is why
localization-delocalization transitions cannot be obtained in this way. Only
perturbational corrections are possible, with the consequence that e.\,g.
the distribution of wave function intensities deviates slightly from the
Porter-Thomas behaviour found in the delocalized regime (see Guhr \etal
1998, Mirlin 2000 for reviews).

In this paper an alternative approach is proposed which avoids the
complication of position-dependent supersymmetry fields. Moreover, it is
even closer to the situation met in experiments, as Efetov's ansatz.

\section{The model}

Let us consider a billiard system with hard walls and statistically
distributed scatterers described by the Hamiltonian

\begin{equation}\label{1}
  H=H_0+V\,,
\end{equation}
where $H_0$  is the operator of kinetic energy with matrix elements

\begin{equation}\label{2}
  \left(H_0\right)_{nm}=E_n^0\delta_{nm}\,
\end{equation}
and $V$ is the potential energy of the scatterers. In Efetov's approach the
potential is assumed to be delta correlated,

\begin{equation}\label{3}
  \left<V(r)V(r')\right>\sim\delta(r-r')\,,
\end{equation}
which gives rise to the above mentioned problems with position-dependent
supersymmetry fields. In this paper the more explicit ansatz

\begin{equation}\label{4}
  V(r)=4\pi\lambda\sum\limits_{l=0}^L\delta(r-r_l)
\end{equation}
is used instead, where the $r_l$   are the positions of the scatterers, and $L$  is its
number. The factor $4\pi$ has been introduced for later convenience. This ansatz dates
back to Lifshitz (1964) and has been applied since then by various authors, among others
Luttinger \etal (1983, 1987).

In the basis of eigenfunctions $\psi_n(r)$ of the billiard without
scatterers the matrix elements of $V(r)$ read

\begin{equation}\label{5}
  V_{nm}=4\pi\lambda\sum\limits_{l=0}^L\psi_n^*(r_l)\psi_m(r_l)\,.
\end{equation}

To simplify the calculations we assume that time-reversal symmetry is
broken, e.\,g. due to the presence of an applied magnetic field.

We are now going to calculate the density of states, averaged over the
positions of the impurities,

\begin{equation}\label{5a}
  \left<\rho(E)\right>=
  -\frac{1}{\pi}\Im\left<\Tr\left(\frac{1}{E_+-H}\right)\right>\,,
\end{equation}

where  $E_+=E+\rmi\epsilon$. Using standard transformations, equation
(\ref{5a}) can be written as

\begin{equation}\label{5b}
  \rho(E)=-\frac{1}{\pi}\left.\frac{\rmd}{\rmd E'}
  \Im\left[Z(E,E')\right]\right|_{E'=E}\,,
\end{equation}
where

\begin{equation}\label{5c}
  Z(E,E')=\left<\frac{\left|E_+'-H\right|}
  {\left|E_+-H\right|}\right>\,.
\end{equation}

$Z$ may be written in terms of an integral over commuting and anti-commuting variables as
(Verbaarschot \etal 1985)

\begin{equation}\label{5d}
\fl Z=\left<\int \rmd[x]\, \exp\left(\rmi
  \sum\limits_{\alpha\beta}
  \left[\left(E_+\delta_{\alpha\beta}-H_{\alpha\beta}\right)x_\alpha^*x_\beta+
  \left(E'_+\delta_{\alpha\beta}-H_{\alpha\beta}\right)\xi_\alpha^*\xi_\beta\right]\right)
  \right>\,,
\end{equation}

where

\begin{equation}\label{7}
  \rmd[x]=\prod\limits_{\alpha=1}^N \rmd x_{\alpha}^*\rmd x_{\alpha}
  \rmd\xi_{\alpha}^* \rmd\xi_{\alpha}
\end{equation}

We adopt the usual convention and use Latin letters for commutating and
Greek ones for anti-commutating variables. In short-hand matrix notation
equation (\ref{5d}) may be written as

\begin{equation}\label{6}
  Z=\left<\int \rmd[x]\, \rm\rme^{\rmi \xf^\dag (\Eb-\Hb)\xf}\right>\,,
\end{equation}

where

\begin{equation}\label{8}
  \xf=\left(x_1,\xi_1,\dots, x_N,\xi_N\right)^T\,,
\end{equation}

\begin{equation}\label{8a}
  \Eb=E\otimes\eb_N=\left(\begin{array}{ccc}
    E & \cdots & 0\\
    \vdots & \ddots &\vdots\\
    0 &\cdots & E \
  \end{array}\right)\,,\qquad
  E=\left
  (\begin{array}{cc}
    E_+ & \cdot \\
    \cdot & E'_+\
  \end{array} \right)\,,
\end{equation}

and

\begin{equation}\label{8b}
  \Hb=\eb\otimes H=\left(\begin{array}{ccc}
    H_{11}\eb & \cdots & H_{1N}\eb\\
    \vdots & \ddots  &\vdots\\
    H_{N1}\eb & \cdots & H_{NN}\eb \
  \end{array}\right)\,.
\end{equation}

In Equations (\ref{8a}) and (\ref{8b})  $\eb_N$  and $\eb$ denote the $N$-
and the two-dimensional unit matrix, respectively. Inserting expression
(\ref{1}) for $H$, equation (\ref{6}) reads

\begin{eqnarray}\label{9}
  Z&=&\int \rmd[x]\, \rme^{\rmi \xf^\dag (\Eb-\Hob) \xf}
  \left<\rme^{-4\pi\rmi\lambda\sum\limits_{l\alpha\beta}
  \psi_\alpha^*(r_l)\psi_\beta(r_l)
  \left(x_\alpha^* x_\beta+\xi_\alpha^* \xi_\beta\right)}\right>\nonumber\\
  &=& \int \rmd[x]\, \rme^{\rmi \xf^\dag (\Eb-\Hob) \xf}M^L\,,
\end{eqnarray}
where

\begin{equation}\label{10}
  M=\left<\rme^{-4\pi\rmi\lambda\sum\limits_{\alpha\beta}
  \psi_\alpha^*(r)\psi_\beta(r) \left(x_\alpha^* x_\beta+\xi_\alpha^* \xi_\beta\right)
  }\right>\,.
\end{equation}

The average in equation (\ref{10}) has to be taken over the positions of the
impurities. But, {\em and this is the central idea of this paper}, instead
of varying over the positions, we may equally well calculate this average by
weighting the expression on the right hand side of the equation with the
joint probability density $p(\psi_{1R}, \psi_{1I}, \dots)$ to find at any
point in the billiard the values $\psi_{1R}$, $\psi_{1I}$, \dots for the
wave function amplitudes. If the billiard without scatterers is chaotic, the
probability density factorizes, $p(\psi_{1R}, \psi_{1I},
\dots)=\prod\limits_\alpha p(\psi_{\alpha R})p( \psi_{\alpha I})$, and real
and imaginary part of the wave functions are Gaussian distributed,

\begin{equation}\label{10a}
  p(\psi_R)=\sqrt{\frac{A}{\pi}}\,\rme^{-A\psi_R^2}\,,\qquad
  p(\psi_I)=\sqrt{\frac{A}{\pi}}\,\rme^{-A\psi_I^2}\,,
\end{equation}
where $A$ is the billiard area. The average (\ref{10}) over the impurity
positions may hence be written as

\begin{equation}\label{11}
  M=\int \prod\limits_\alpha \left[\rmd\psi_{\alpha R}\,\rmd\psi_{\alpha I}\,
  p(\psi_{\alpha R})p(\psi_{\alpha I})\right]\,
  \rme^{-4\pi\rmi\lambda\sum\limits_{\alpha\beta}
  \psi_\alpha^*\psi_\beta \left(x_\alpha^* x_\beta+\xi_\alpha^*
  \xi_\beta\right)}\,.
\end{equation}

With the weight functions (\ref{10a}) the integrations are easily performed
yielding

\begin{equation}\label{12}
  M=\frac{1}{\left|\eb_N+\rmi\frac{4\pi\lambda}{A}X\right|}\,,
\end{equation}
where $X$ is the $N\times N$ matrix  with the elements

\begin{equation}\label{13}
  X_{\alpha\beta}= x_\alpha^*x_\beta+\xi_\alpha^*\xi_\beta\,.
\end{equation}

According to the Weyl formula the mean density of states in two-dimensional
billiards is given by $\left<\rho\right> =A/4\pi$. Following the usual
practice we normalize this quantity to one, and omit the factor $4\pi/A$ in
the following. The determinant (\ref{12}) is now transformed by means of the
relation

\begin{equation}\label{14}
  |\eb_N+AB|=|\eb_M+BA|\,
\end{equation}
holding for arbitrary $N\times M$ matrices $A$, and $M\times N$ matrices
$B$. This follows in a straightforward manner from the relation
$|M|=\exp[\Tr(\ln M)]$. It is not necessary that the matrices are quadratic
providing us with an efficient tool to reduce the rank of determinants.
Applied to equation (\ref{12}) relation (\ref{14}) yields

\begin{equation}\label{15}
  M=\frac{1}{\left|\eb+\rmi\lambda\hat{X}\right|}\,
\end{equation}
where $\hat{X}$   is the $2\times 2$ supermatrix

\begin{equation}\label{16}
  \hat{X}=\left(\begin{array}{cc}
    \sum\limits_\alpha x_\alpha x_\alpha^* & \sum\limits_\alpha x_\alpha \xi_\alpha^* \\
    \sum\limits_\alpha \xi_\alpha x_\alpha^* & \sum\limits_\alpha \xi_\alpha \xi_\alpha^*\
  \end{array}\right)\,.
\end{equation}

We have thus arrived at the intermediate result

\begin{equation}\label{17}
  Z= \int \rmd[x]\, \rme^{\rmi \xf^\dag(\Eb-\Hob) \xf}
  \left|\eb+\rmi\lambda\hat{X}\right|^{-L}
\end{equation}

Whenever there are super matrices involved,  determinants and traces have to be
interpreted as super determinants and super traces, respectively, where we shall use the
convention of Verbaarschot \etal 1985.

It is instructive to consider the small $\lambda$ limit of expression
(\ref{17}). The determinant may be expanded as

\begin{eqnarray}\label{18}
  \left|\eb+\rmi\lambda\hat{X}\right|^{-L}&=&
  \exp\left[-L\Tr\ln\left(1+\rmi\lambda\hat{X}\right)\right]\nonumber\\
  &=&\exp\left[-\rmi L\lambda\Tr\hat{X}
  -\frac{L\lambda^2}{2}\Tr\hat{X}^2+\cdots\right]\,.
\end{eqnarray}

Stopping at the quadratic term, equation (\ref{17}) reads

\begin{equation}\label{19}
  Z= \int \rmd[x]\, \rme^{\left[\rmi \xf^\dag(\Eb-\Hob-L\lambda\eb_N) \xf
  -\frac{L\lambda^2}{2}\Tr \hat{X}^2\right]}\,,
\end{equation}
where $\Tr \hat{X}=\xf^\dag \xf$ was used.  This is exactly the expression
obtained for the ensemble average of the Hamiltonian

\begin{equation}\label{20}
  H=H_0+L\lambda\eb_N+H_1\,.
\end{equation}
where the matrix elements of $H_1$  are Gaussian random variables with variance
$\left<H_1^2\right>=L\lambda^2$. We thus can note already at this early stage that in the
small $\lambda$ limit random matrix results will be recovered.

\section{The $x$ integrations}\label{sec3}

The usual approach to perform integrations of type (\ref{19}) is a Hubbard-Stratonovich
transformation to remove the $\Tr\hat{X}^2$    term in the exponent, depending on the
integration variables in the fourth order. As a result the $x$ integrations reduce to
simple Fresnel integrals which are trivially solved.

For the integral (\ref{17}) a Hubbard-Stratonovich transformation is not
possible. One way to proceed further is to write again the determinant in
terms of a superintegral,

\begin{equation}\label{21}
  \frac{1}{\left|\eb+\rmi\lambda\hat{X}\right|}=
  \int \rmd[y] \rme^{-{\bf y}^\dag \left(\eb+\rmi\lambda\hat{X}\right){\bf y}}\,.
\end{equation}

We need $L$ replicas of this integral since the determinant enters in the
$L$th power, leading to the introduction of $4L$ new integration variables.
The $x$ integrations can then be performed in the usual manner.

To avoid the introduction of such a large number of new integration
variables, we apply another approach. Let us consider the integral

\begin{equation}\label{22}
  I_L(A)=\int \rmd[t] |T|^L \rme^{-\Tr(AT)}\,,
\end{equation}
 where

\begin{equation}\label{22a}
A=\left(\begin{array}{cc}
    a & \alpha^* \\
    \alpha & \bar{a} \
  \end{array}\right)\,,\qquad
  T=\left(\begin{array}{cc}
    t & \tau^* \\
    \tau & \bar{t} \
  \end{array}\right)
\end{equation}
are supermatrices of rank $2$.

Because of the basis independence of trace and determinant it is always
possible to chose the $T$ variables such that $A$ is diagonal, i.\,e.\,
$\alpha=\alpha^*= 0$. Equation (\ref{22}) then reads

\begin{equation}\label{23}
  I_L(A)=\int \rmd t \,\rmd\bar{t}\,\rmd\tau^*\,\rmd\tau
  \left|\begin{array}{cc}
    t & \tau^* \\
    \tau & \bar{t} \
  \end{array}\right|^L \rme^{-(at-\bar{a}\bar{t})}\,,
\end{equation}

Introducing new integration variables $s=at$, $\bar{s}=\bar{a}\bar{t}$,
$\sigma^*=a\tau^*$, $\sigma=\bar{a}\tau$ we obtain

\begin{eqnarray}\label{24}
  I_L(A)&=&\int \frac{\rmd s}{|a|}
  \,\frac{\rmd\bar{s}}{|\bar{a}|}\,|a|\,d\sigma^*\,|\bar{a}|\,d\sigma
  \left|\begin{array}{cc}
    a^{-1} s\, & a^{-1}\sigma^* \\
    \bar{a}^{-1}\sigma\, & \bar{a}^{-1}\bar{s} \
  \end{array}\right|^L \rme^{-\Tr(s-\bar{s})}\\\nonumber
&=& \left(\frac{|\bar{a}|}{|a|}\right)^L I_L\,,
\end{eqnarray}
where
\begin{equation}\label{25}
  I_L=\int \rmd[t] |T|^L \rme^{-\Tr(T)}\,,
\end{equation}
or
\begin{equation}\label{26}
  \left|A\right|^{-L}=I_L(A)/I_L\,.
\end{equation}

Because of the basis independence of this expression the latter result holds for
arbitrary supermatrices $A$, not necessarily diagonal.

This is an alternative to equation (\ref{21}) to express the power of a
determinant as a superintegral, avoiding the need to introduce $L$ replicas.
The question arises, how the paths of integration are to be chosen to make
the integral (\ref{22}) well-defined. From equation (\ref{22a}) we have

\begin{equation}\label{28}
  \Tr\, T= t-\bar{t}\,,\qquad
  |T|=\left.\left(t-\frac{\tau^*\tau}{\bar{t}}\right)\right/\bar{t}\,.
\end{equation}

Shifting the variable $t$ by $\tau^*\tau/\bar{t}$, integral (\ref{25}) reads

\begin{equation}\label{29}
  I_L= \int \rmd t\,\rmd\bar{t}\,\rmd\tau^*\,\rmd\tau\,
  \left|\frac{t}{\bar{t}}\right|^L \rme^{-\left(t+\tau^*\tau/\bar{t}-\bar{t}\right)}\,.
\end{equation}

The integration over the antisymmetric variables is straightforward, and we
are left with the $t$, $\bar{t}$  integrations,

\begin{equation}\label{30}
  I_L=\frac{1}{2\pi}\int \rmd t\,\rmd\bar{t}\,\frac{t^L}{\bar{t}^{L+1}}\,
  \rme^{-(t-\bar{t})}\,.
\end{equation}

Let us assume for a moment that $L$ is non-integer. Then we may define an
integration path starting at $\rme^{\rmi\phi}\infty$, encircling the origin
counterclockwise, and returning to $\rme^{\rmi\phi}\infty$. The phase angle
$\phi$  has to be chosen in a way that the integral is well-defined. We end
thus with the well-known integral representation for the reciprocal Gamma
function, both for the $t$ and the $\bar{t}$ integration, with the result

\begin{eqnarray}\label{31}
  I_L&=&\frac{2\pi}{\Gamma(L+1)\Gamma(-L)}\,\rme^{\rmi\pi L}=
  2\sin\pi (L+1)\, \rme^{\rmi\pi L}\\\nonumber
  &=& \rmi\left( \rme^{2\pi\rmi L}-1\right)\,.
\end{eqnarray}

Equation (\ref{26}) thus is well-defined for non-integer $L$ if the paths of
integration are chosen as described above. For integer $L$ the expression on
the right hand side is not defined, but it is easily seen that the limit
(non-integer $L$) $\to$ (integer $L$) exists and gives

\begin{equation}\label{32}
  \left|A\right|^{-L}=\frac{1}{\rmi}\int\limits_0^{\rme^{\rmi\phi}\infty}\rmd t
  \,\oint  \rmd\bar{t}\,\rmd\tau^*\,\rmd\tau\,\left|T\right|^L\rme^{-\Tr (AT)}\,,
\end{equation}
where the $\bar{t}$ integration is performed counterclockwise on a circle
about the origin. Equation (\ref{32}) holds for all natural numbers $L$.

Applied to equation (\ref{17}), we have

\begin{equation}\label{33}
  Z= \frac{1}{I_L}\int \rmd[x]\,\rmd[t]\, \rme^{\rmi \xf^\dag(\Eb-\Hob) \xf}
 |T|^L \rme^{-\Tr \left[T\left(\eb+\rmi\lambda\hat{X}\right)\right]}
\end{equation}

Now the $x$ integrations can be performed, using definition (\ref{16}),

\begin{eqnarray}\label{34}
  \int \rmd[x]\,
  \rme^{\rmi \left[\xf^\dag(\Eb-\Hob)\xf-\Tr(\lambda T\hat{X})\right]}
  &=& \int \rmd[x]\,
  \rme^{\rmi \xf^\dag(\Eb-\Hob-\lambda T\eb_N)\xf}\nonumber\\
  &=&\prod_\alpha\frac{1}{\left|E-E_a^0\eb- \lambda T\right|}\,,
\end{eqnarray}
whence follows

\begin{equation}\label{35}
  Z= \frac{1}{I_L}\int \rmd[t]\, |T|^L \rme^{-\Tr T}
  \prod_\alpha\frac{1}{\left|E-E_a^0\eb-\lambda T\right|}\,.
\end{equation}

This may alternatively be written as

\begin{equation}\label{36}
  Z= \frac{1}{I_L}\int \rmd[t]\, \rme^{-\Tr\left[F(T)\right]}\,,
\end{equation}
where

\begin{equation}\label{37}
  F(T)= T-L\ln T+\sum_\alpha \ln\left(E-E_a^0{\bf
  1}-\lambda T\right)\,.
\end{equation}

Equations (\ref{35}) to (\ref{37}) constitute our next intermediate result. They allow to
calculate the averaged density of states of a billiard with randomly distributed
scatterers in terms of the eigenenergies of the billiard without scatterers. All
integrals can be solved exactly by means of the residuum method. In the remaining step
the limit $N\to\infty$ has to be performed. The occurring infinite products diverge as a
consequence of the delta-like singularities in the potential. But the divergencies can be
handled in a standard way by a renormalization of the coupling constant (Albeverio and
\v{S}eba 1991). See also Bogomolny \etal 2001, where the situation of a single scatterer
in a rectangular billiard is studied.

From microwave experiments it is known, but for systems with time-reversal symmetry only,
that in billiards with randomly distributed scatterers the wave functions are localized
at low energies, but  become delocalized at sufficiently high energies (Kudrolli \etal
1995, St\"ockmann \etal 2001). A calculation of the averaged density of states as a
function of energy from equation (\ref{35}), and  of two-point correlation function,
inverse participation ratio etc. from its generalization should thus exhibit clear
fingerprints of the localization-delocalization transition.

This will be the program for future works. For the moment let us proceed
along a more convenient route by taking $N$ fixed and finite, and by
performing a second average over the shape of the billiard.

\section{The average over the billiard shape}\label{sec4}

According to a conjecture of Bohigas, Giannoni, Schmit (1984) the spectrum of a billiard
with broken time-reversal symmetry should obey the same statistical features as the
spectrum of a random matrix taken from the Gaussian Unitary Ensemble (GUE). Taken this
for granted we may replace $H_0$ in equation (\ref{33}) by a GUE matrix and perform a
Gaussian average over the matrix elements to obtain the average over the billiard shape.
(Up to now $H_0$ had been assumed to be diagonal, but because of the basis invariance of
the expression we may take any other basis as well; it is much easier to perform the
average over the matrix elements than over the eigenvalues.)

The Gaussian average over the matrix elements is trivial and yields

\begin{eqnarray}\label{38}
  \left<\rme^{-\rmi\xf^\dag \Hob\xf}\right> &=&
  \left< \rme^{-\rmi\sum\limits_{\alpha\beta}
   \left(H_0\right)_{\alpha\beta}
   (x_\alpha^*x_\beta+\xi_\alpha^*\xi_\beta)}\right>\\\nonumber
   &=& \left< \rme^{-\frac{1}{2}\left<\left(H_0\right)^2\right>
   \sum\limits_{\alpha\beta}
   (x_\alpha^*x_\beta+\xi_\alpha^*\xi_\beta)(x_\beta^*x_\alpha+\xi_\beta^*\xi_\alpha)}
   \right>\\\nonumber
   &=& e^{-\frac{N}{2\pi^2}\Tr\left(\hat{X}\right)^2}\,.
\end{eqnarray}

Following  common practise again we have shifted the average energy to zero, and have
applied the normalization $\left<\left(H_0\right)^2\right>=N/\pi^2$ yielding a mean
density of states of one at $E=0$ (Verbaarschot \etal 1985). After a subsequent
Hubbard-Stratonovich transformation equation (\ref{38}) reads

\begin{equation}\label{39}
   \left<\rme^{-\rmi\xf^\dag \Hob\xf}\right> =
   \int \rmd[y]\,\rme^{-\frac{\pi^2}{2N}\Tr Y^2-\rmi\Tr\left(\hat{X}Y\right)}\,,
\end{equation}
where

\begin{equation}\label{40}
  Y=\left(\begin{array}{cc}
    y & \eta^* \\
    \eta & \bar{y} \
  \end{array}\right)\,.
\end{equation}

To make expression (\ref{39}) well-defined, the $y$ integration has to be
performed from $-\infty$ to $\infty$, and the $\bar{y}$ integration from
$-\rmi\infty$ to $\rmi\infty$. Inserting expression (\ref{39}) into equation
(\ref{33}) we get as the result of the shape averaging

\begin{eqnarray}\label{41}
  \left<Z\right>&=&\frac{1}{I_L}\int \rmd[x]\,\rmd[t]\,\rmd[y]\,\left|T\right|^L
  \rme^{-\Tr T}\,\rme^{-\frac{\pi^2}{2N}\Tr Y^2}\,
  \rme^{\rmi\xf^\dag(\Eb-\lambda
  T\eb_N)\xf}\,\rme^{-\rmi\Tr\left(\hat{X}Y\right)}\\\nonumber
 &=&\frac{1}{I_L}\int \rmd[t]\,\rmd[y]\,\left|T\right|^L
  \rme^{-\Tr T}\,\rme^{-\frac{\pi^2}{2N}\Tr Y^2}
  \int \rmd[x] \rme^{\rmi\xf^\dag(E-\lambda T-Y)\eb_N\xf}\,,
\end{eqnarray}
where we have used $\Tr(\hat{X}Y)=\xf^\dag Y\eb_N\xf$. The $X$ integrations
are straightforward and yield

\begin{equation}\label{42}
   \left<Z\right>= \frac{1}{I_L}\int \rmd[t]\,\rmd[y]\,\left|T\right|^L
  \rme^{-\Tr T}\,\rme^{-\frac{\pi^2}{2N}\Tr Y^2}
  \frac{1}{|E-\lambda T-Y|^N}\,.
\end{equation}

Again we apply expression (\ref{26}) to rewrite the determinant,

\begin{equation}\label{43}
  \frac{1}{|E-\lambda T-Y|^N}=\frac{1}{I_N}\int \rmd[s]\,\left|S\right|^N\rme^{\Tr S(E- \lambda T-Y)}\,,
\end{equation}
where

\begin{equation}\label{44}
    S=\left(\begin{array}{cc}
    s & \sigma^* \\
    \sigma & \bar{s}
  \end{array}\right)\,.
\end{equation}

In addition, we replace $T$ by $NT$, and $Y$ by $NY$ and obtain

\begin{eqnarray}\label{45}
\fl   \left<Z\right>&=& \frac{1}{I_L I_N}\int
\rmd[t]\,\rmd[y]\,\rmd[s]\,\left|T\right|^L
  \rme^{-N\Tr T}\,\rme^{-N\frac{\pi^2}{2}\Tr Y^2}\,\left|S\right|^N\,
  \rme^{\Tr S(E-\lambda NT-NY)}\\\nonumber
\fl  &=& \frac{1}{I_N}\int \rmd[s]\,\left|S\right|^N\,\rme^{\Tr(SE)}\,
  \frac{1}{I_L}\int \rmd[t]\left|T\right|^L\,\rme^{-N\Tr T({\eb}+\lambda S)}
  \int \rmd[y]\,\rme^{-N\left(\frac{\pi^2}{2}\Tr Y^2+\Tr(SY)\right)}\,.
\end{eqnarray}

The $T$ and  the $Y$ integrations can now be performed with the result

\begin{equation}\label{46}
   \left<Z\right> = \frac{1}{I_N}\int \rmd[s]\, \rme^{\Tr(SE)}\frac{\left|S\right|^N}
   {\left|{\eb}+\lambda S\right|^L}\,\rme^{\frac{N}{2\pi^2}\Tr S^2}\,.
\end{equation}

Equation (\ref{46}) is the main result of this paper. It is surprisingly
simple and allows an easy calculation of the density of states of  the
billiard with randomly scatterers, doubly averaged over disorder and shape
of the billiard. We only have to perform the remaining integrations over 4
commuting and anti-commuting variables.

\section{The density of states}

The calculation of the integral is easiest, if we transform the matrix $S$
into a diagonal matrix via

\begin{equation}\label{47}
\fl    S=\left(\begin{array}{cc}
    s & \sigma^* \\
    \sigma & \bar{s}
  \end{array}\right) =
  \left(\begin{array}{cc}
    \sqrt{1+\beta\gamma} & -\beta \\
    -\gamma & \sqrt{1+\gamma\beta}
  \end{array}\right)
  \left(\begin{array}{cc}
    s_B & \cdot \\
    \cdot & s_F
  \end{array}\right)
  \left(\begin{array}{cc}
    \sqrt{1+\beta\gamma} & \beta \\
    \gamma & \sqrt{1+\gamma\beta}
  \end{array}\right)\,.
\end{equation}

After performing the matrix multiplications we have

\begin{equation}\label{48}
\begin{array}{rclrcl}
    s&=&s_B+\beta\gamma\left(s_B-s_F\right)\,,& \qquad \sigma^*&=&\beta\left(s_B-s_F\right)\,,\\
        \sigma&=&-\gamma\left(s_B-s_F\right)\,,& \qquad
    \bar{s}&=&s_F+\beta\gamma\left(s_B-s_F\right)\,,
\end{array}
\end{equation}
whence follows for the volume element

\begin{equation}\label{49}
  \rmd[s]= -\frac{\rmd s_B\,\rmd s_F\,\rmd\beta\,\rmd\gamma}{\left(s_B-s_F\right)^2}\,.
\end{equation}

Remembering that the $s_B$ integration is from 0 to $\rme^{\rmi\phi}\infty$
with a suitably chosen phase angle $\phi$, and that the $s_F$ integration is
along a circle about the origin (see equation (\ref{32})), we obtain from
equation (\ref{46})

\begin{eqnarray}\label{50}
  \left<Z\right>&=& -\frac{1}{\rmi}\int\limits_0^{\rme^{\rmi\phi}\infty}\oint
  \frac{\rmd s_B\,\rmd s_F\,\rmd\beta\,\rmd\gamma}{\left(s_B-s_F\right)^2}
  \rme^{E_+\left[s_B+\beta\gamma\left(s_B-s_F\right)\right]-
  E_+'\left[s_F+\beta\gamma\left(s_B-s_F\right)\right]}\\\nonumber
  &&\times \left|\frac{s_B}{s_F}\right|^N
  \left|\frac{1+\lambda s_F}{1+\lambda s_B}\right|^L
  \rme^{\frac{N}{2\pi^2}\left(s_B^2-s_F^2\right)}\,.
\end{eqnarray}

The value for the phase angle can be inferred from equation (\ref{43}):
since $t$ and $y$ are real, and $E_+$ has an infinitesimally small positive
imaginary part, the integration has to be performed from 0 to $\rmi\infty$.

The integral over the antisymmetric variables is easily done and yields

\begin{eqnarray}\label{51}
  \left<Z\right>&=& \frac{1}{2\pi\rmi}\int\limits_0^{\rmi\infty}\rmd s_B\,
  \oint \rmd s_F
  \frac{E_+-E_+'}{s_B-s_F}\,
  \rme^{E_+s_B-E_+'s_F}\\\nonumber
  &&\times \left|\frac{s_B}{s_F}\right|^N
  \left|\frac{1+\lambda s_F}{1+\lambda s_B}\right|^L\,
  \rme^{\frac{N}{2\pi^2}\left(s_B^2-s_F^2\right)}\,.
\end{eqnarray}

It follows for the mean density of states (see equation (\ref{5b}))

\begin{eqnarray}\label{52}
\fl \left<\rho(E)\right>&=&-\frac{1}{\pi}\left.\Im\frac{\rmd Z_1}{\rmd
E'}\right|_{E'=E}\\\nonumber \fl
&=&\frac{1}{\pi}\Im\frac{1}{2\pi\rmi}\int\limits_0^{\rmi\infty}\rmd s_B\,
\oint \rmd s_F
  \frac{\rme^{E(s_B-s_F)}}{s_B-s_F}\,\left|\frac{s_B}{s_F}\right|^N
  \left|\frac{1+\lambda s_F}{1+\lambda s_B}\right|^L\,
  \rme^{\frac{N}{2\pi^2}\left(s_B^2-s_F^2\right)}\,.
\end{eqnarray}

Differentiating with respect to $E$, we have after some straightforward
transformations

\begin{equation}\label{53}
  \left<\rho'(E)\right>=\frac{\pi^2}{2N}I_{NL}(\epsilon,\alpha)
  \bar{I}_{(N-1)L}(\epsilon,\alpha)\,,
\end{equation}
where

\begin{equation}\label{54}
  \epsilon=\frac{\pi}{\sqrt{2N}}E\,,\qquad
  \alpha=\frac{\sqrt{N/2}}{\pi\lambda}\,,
\end{equation}
and

\begin{equation}\label{55}
  I_{NL}(\epsilon,\alpha)=\frac{1}{\pi\rmi}
  \int\limits_{-\rmi\infty}^{\rmi\infty} \rmd x\,
  \rme^{2\epsilon x}\frac{(2x)^N}{(x+\alpha)^L}\,\rme^{x^2}\,,
\end{equation}
\begin{equation}\label{55a}
    \bar{I}_{NL}(\epsilon,\alpha)=\frac{1}{\pi\rmi}\oint \rmd y\,\rme^{-2\epsilon y}\,
  \frac{(y+\alpha)^L}{(2y)^{N+1}}\,\rme^{-y^2}\,.
\end{equation}

From the definitions we immediately obtain the recursion relations

\begin{equation}\label{56}
\begin{array}{crlrcl}
 I_{NL}'&=&I_{(N+1)L}\,,\qquad
 &\left(I_{NL}\rme^{2\epsilon\alpha}\right)'&=&I_{N(L-1)}\rme^{2\epsilon\alpha}\\
  \bar{I}_{NL}'&=&-\bar{I}_{(N-1)L}\,,\qquad &
  \left(\bar{I}_{NL}\rme^{-2\epsilon\alpha}\right)'&=&-\bar{I}_{N(L+1)}\rme^{-2\epsilon\alpha}\,,
\end{array}
  \end{equation}
where the prime denotes differentiation with respect to $\epsilon$.

For $L=0$ we get in particular

\begin{eqnarray}\label{57}
    I_{N0}(\epsilon,\alpha)&=&\frac{(-1)^N}{\sqrt{\pi}}\,
    \rme^{-\epsilon^2}H_N(\epsilon)\,,\\
  \bar{I}_{N0}(\epsilon,\alpha)&=&\frac{(-1)^N}{2^N N!}\,
  H_N\left(\epsilon\right)\,,
\end{eqnarray}
where integral representations of the Hermite polynomials have been used
(see e.\,g. Magnus \etal 1966). Using the recursion relations we now
calculate $\left<\rho(E)\right>$ from equation (\ref{53}) by repeated
partial integration with the result

\begin{equation}\label{58}
  \left<\rho(E)\right>=\frac{\pi}{\sqrt{2N}}\sum\limits_{k=0}^{N-1}
  I_{kL}(\epsilon,\alpha)\bar{I}_{kL}(\epsilon,\alpha)\,.
\end{equation}

We have thus obtained a closed expression for the averaged density of states
for arbitrary values of $N$ and $L$. It is an easy matter to show, again
using the recursion relations (\ref{56}), that $\int\limits_{-\infty}^\infty
\left<\rho(E)\right>\,\rmd E=N$, as it should be.

For $L=0$ equation (\ref{58}) reduces to

\begin{equation}\label{59}
  \left<\rho(E)\right>=\frac{\pi}{\sqrt{2N}}\sum\limits_{k=0}^{N-1}
  \left[\psi_k(\epsilon)\right]^2\,,
\end{equation}

where

\begin{equation}\label{60}
  \psi_k(x)=\frac{1}{\left(2^kk!\sqrt{\pi}\right)^{1/2}}H_k(x)\,
  \rme^{-x^2/2}
\end{equation}
is an harmonic oscillator eigenfunction. This is identical with the well-known exact
expression for the density of states of the Gaussian unitary ensemble, which in the limit
of large $N$ reduces to Wigner's semicircle law (Mehta 1991).

\section{The strong coupling limit}

Using the recursion relations (\ref{56}) $I_{NL}$ and $\bar{I}_{NL}$ can be
calculated from $I_{N0}$ and $\bar{I}_{N0}$ by repeated integration or
differentiation, respectively. Since all integrations can be performed
analytically, we have got an exact representation for the density of states
for arbitrary $L$. Though this may be helpful for small values of $L$, it is
not very useful for practical purposes, since one is usually interested in
the limit $N$, $L\to\infty$ while the ratio $l=L/N$ remains finite.

In such a situation it suggests itself to solve the integrals (\ref{55}) and
(\ref{55a}) with help of saddle point techniques. This leads to a cubic
saddle-point equation which still can be solved exactly using Cardano's
formula. The resulting equations are not very elucidating, however.
Therefore we proceed in another direction and restrict the following
discussion to the strong coupling limit $\lambda\gg 1$, or $\alpha\ll 1$. In
the discussion we have to discriminate between the two situations $N>L$ and
$N<L$.

(i) $N>L$

For this case we may replace $(x+\alpha)^L$ and $(y+\alpha)^L$ in the
integrands by $x^L$ and $y^L$, respectively, to obtain

\begin{equation}\label{61}
  I_{NL}(\epsilon,\alpha)=\frac{2^L}{\pi\rmi}
  \int\limits_{-\rmi\infty}^{\rmi\infty} \rmd x\,
  \rme^{2\epsilon x}(2x)^{N-L}\,\rme^{x^2}\,,
\end{equation}
\begin{equation}\label{61a}
    \bar{I}_{NL}(\epsilon,\alpha)=\frac{2^{-L}}{\pi\rmi}\oint \rmd y\,\rme^{-2\epsilon y}\,
  (2y)^{-(N-L)}\,\rme^{-y^2}\,,
\end{equation}

In the strong coupling limit the averaged density of states for a billiard
system with $N$ levels taken into account and $L$ randomly distributed
scatterers is thus the same as for a system with $N-L$ levels, and no
scatterer at all. We are again in the random matrix regime.

Remember that already in the beginning we observed that the spectra of
billiards with randomly distributed scatterers show random matrix behaviour,
but at that point we considered the {\em weak} coupling limit $\lambda\ll 1$
( see the discussion following equation (\ref{18})).

We thus can note that for $N>L$ both  in the weak and the strong coupling
limit the averaged density of states shows random matrix behaviour.

(ii) $N<L$

Now we cannot replace any longer $(x+\alpha)^L$ and $(y+\alpha)^L$ in the
integrands in equations (\ref{55}) and (\ref{55a}) by $x^L$ and $y^L$, since
in this limit the integral for $I_{NL}$ diverges, and that for
$\bar{I}_{NL}$ gives zero. In the limit $\alpha\ll 1$, on the other hand,
the main contributions to the integrals come from regions $x\ll 1$ and $y\ll
1$, where the Gaussian cut-offs are not yet relevant. We may therefore
replace $\rme^{x^2}$ and $\rme^{-y^2}$ by one,  and solve the integrals by
means of the residuum method with the result

\begin{eqnarray}\label{62}
  I_{NL}(\epsilon,\alpha) &=& 2^{N+1}\frac{\Theta(\epsilon)}{(L-1)!}
  \left(\frac{\rmd}{\rmd x}\right)^{L-1}
  \left.\left(\rme^{2\epsilon x}x^N\right)\right|_{x=-\alpha}\\\nonumber
  &=& 2^{N+1}\alpha^{N+1-L}\Theta(z)(-1)^{N+1-L}e^{-z}L_{L-1}^{(N-L+1)}(z)\\\nonumber
  &=& 2^{N+1}\alpha^{N+1-L}\Theta(z)\frac{N!}{(L-1)!}e^{-z}z^{L-N-1}L_N^{(L-N-1)}(z)\,,
 \end{eqnarray}

\begin{eqnarray}\label{63}
 \bar{I}_{NL}(z,\alpha)&=& 2^{-N}\frac{1}{N!}\left(\frac{\rmd}{\rmd y}\right)^N
  \left.\left(e^{-2\epsilon y}(y+\alpha)^L\right)\right|_{y=0}\\\nonumber
  &=& 2^{-N}\alpha^{L-N} L_{N}^{(L-N)}(z)\,,
\end{eqnarray}
where

\begin{equation}\label{63a}
  z=2\epsilon\alpha=E/\lambda\,.
\end{equation}

$\Theta(z)$ is the Heaviside step function, and $L_n^{(\alpha)}(z)$ is a
generalized Laguerre polynomial. (There are two conventions for the Laguerre
polynomials found in literature, differing in the normalization. In this
paper the definition of Magnus \etal 1966 is adopted, where
$L_n^{(\alpha)}(0)={n+\alpha \choose n}$.) It follows from equation
(\ref{58}) for the density of states

\begin{equation}\label{64}
  \left<\rho(E)\right>=\frac{1}{\lambda(L-1)!}\Theta(z)\,e^{-z}
  \sum\limits_{k=0}^{N-1} k!\,
  z^{L-k-1}L_k^{(L-k-1)}(z)\,L_k^{(L-k)}(z)\,.
\end{equation}

Equation (\ref{64}) simplifies considerably in the limit $L\to\infty$,
$N\to\infty$, with $L/N$ remaining finite. Inserting expressions (\ref{62})
and (\ref{63}) for $I_{NL}(z,\alpha)$ and $\bar{I}_{NL}(z,\alpha)$,
respectively, into equation (\ref{53}) we obtain

\begin{equation}\label{65}
   \left<\rho'(E)\right>=\frac{1}{\lambda^2}\,\frac{N!}{(L-1)!}\,\Theta(z)\,e^{-z}
   z^{L-N-1}L_N^{(L-N-1)}(z)\,L_{(N-1)}^{(L-N+1)}(z)\,.
\end{equation}

In terms of the function

\begin{equation}\label{66}
  y_n^{(\alpha)}(z)=e^{-\frac{z}{2}}z^{\frac{\alpha+1}{2}}L_n^{(\alpha)}(z)\,,
\end{equation}
equation (\ref{65}) may be written as

\begin{equation}\label{67}
  \left<\rho'(E)\right>=\frac{1}{\lambda^2}\,\frac{N!}{(L-1)!}\,\Theta(z)
  z^{-2}\,y_N^{(L-N-1)}(z)\,y_{N-1}^{(L-N+1)}(z)\,.
\end{equation}

The $y_n^{(\alpha)}$ obey the differential equation

\begin{equation}\label{68}
  y''+\left(\frac{2n+\alpha+1}{2z}-\frac{1}{4}+\frac{1-\alpha^2}{4z^2}\right)y=0\,.
\end{equation}

Equation (\ref{68}) is easily identified as the radial Schr\"odinger
equation of the hydrogen atom, where $y_n^{(\alpha)}(z)/z$ is the radial
part of the wave function. This suggests an approximation of
$y_n^{(\alpha)}$ by means of the WKB method. In the present context it is
sufficient to consider the solution in the classically allowed region. For
this regime the WKB approximation yields (see e.\,g. section 9.3 of Morse
and Feshbach 1953)

\begin{equation}\label{69}
  y_n^{(\alpha)}(z)=\frac{y_0}{\sqrt{q}}
  \cos\left(\,\int\limits_{z_0}^z q\,\rmd z-\frac{\pi}{4}\right)\,,
\end{equation}
where
\begin{equation}\label{70}
  q=\sqrt{\frac{2n+\alpha+1}{2z}-\frac{1}{4}-\frac{\alpha^2}{4z^2}}\,,
\end{equation}
and $z_0$, $z_1$ are the classical turning points given by

\begin{equation}\label{71}
  z_{0/1}=2n+\alpha+1\pm\sqrt{(2n+1)(2n+2\alpha+1)}\,.
\end{equation}

The replacement of $1-\alpha^2$ by $-\alpha^2$ in going from equation (\ref{68}) to
equation (\ref{70}) corrects for the singularity of the potential at $z=0$, see the
discussion in Morse und Feshbach 1953. (The same technique can be applied to derive the
semi-circle law in a simple way from the exact expression (\ref{59}), see chapter 3.2.3
of St\"ockmann 1999; the procedure is more or less an elaboration of an idea developed in
appendix A.9 of the book of Mehta (1991).)

Inserting approximation (\ref{69}) into equation (\ref{67}) we end up with

\begin{equation}\label{72}
  \left<\rho(E)\right>=\frac{1}{\lambda\pi}\frac{1}{2z}\,
  \sqrt{4LN-(z-L-N)^2},\qquad z=E/\lambda\,,
\end{equation}

where the classical turning points are given by

\begin{equation}\label{74}
  z_{0/1}=L+N\pm2\sqrt{LN}\,.
\end{equation}

Details of the derivation can be found in the appendix. The density of
states thus changes dramatically if $L$ surpasses $N$. For $L<N$ Wigner's
semicircle law is found, and the eigenenergies are distributed between
$-2\pi/N$ and $2\pi/N$. For $L>N$ on the other hand only positive
eigenvalues are found, if $\lambda$ is positive, in an energy window limited
by $\lambda z_0$ and $\lambda z_1$.

\section{Discussion}

We have obtained a surprisingly simple expression for the averaged density of states of a
billiard with randomly distributed scatterers. The central ingredient was the idea to
substitute the average over the scatterer positions in equation (\ref{10}) by an weighted
average, with the wave function amplitude probability density as the weight function. It
was argued that both averages are equivalent. In view of the central importance of this
procedure it seems appropriate to discuss the limitations of the approach.

(i) First, the impurities are considered as uncorrelated. In particular, it
is not excluded that two impurities occupy the same site.

(ii) Second, wave functions belonging to different eigenvalues are considered as
uncorrelated. This may pose a problem, since it is known from semi-classical quantum
mechanics that there are correlations on energy scales of the order of $\hbar/T$, where
$T$ is the length of the shortest periodic orbit (see Gutzwiller 1990 for a review). On
the other hand, these correlations vanish in the semi-classical limit on energy scales of
the mean level spacing. It therefore seems legitimate to neglect correlations between
different wave functions.

No problem, on the other hand, arises from the fact that there are spatial correlations
for individual wave functions, as is well-known from the works of Berry (1977) and Fal'ko
and Efetov (199). Since only the weight of the wave function amplitudes enters equation
(\ref{11}), spatial correlations are completely irrelevant.

The approximation performed in section \ref{sec4} by substituting the billiard by a
random matrix of finite rank is of another type. It has been applied to obtain a simple
tractable model, but  by this second step we have reduced our system to a mere caricature
of a real billiard system. In particular the information on the dimension of the
billiard, which obviously is an important quantity for questions of localization and
delocalization, is lost. (The information on the dimension is still present in equation
(\ref{35}), namely in the spectrum of the empty billiard which depends on the dimension
via the mean density of states.)

This is why in the moment a comparison with literature results is not possible. In
particular the work of Luttinger and Tao (1983) has to be mentioned in this respect, who
calculated the density of states for the billiard with randomly distributed scatterers in
the low energy limit. For a more detailed consideration of their results, we would have
to go back to equations (\ref{35}) to (\ref{37}), perform the limit $N\to\infty$, and
calculate the density of states for the true billiard system, and not a random matrix
substitute only.

But the present results suggest that already in our toy model there is a
localization-delocalization transition at $L=N$. For $L<N$ we are in the
regime of delocalized wave functions obeying random-matrix behaviour. For
$L>N$, on the other hand the wave functions become localized, giving rise to
a completely changed density of states. In this respect the rank $N$ of the
matrix seems to take the role of the energy in the real billiard system.

For the moment, however, this conclusion must be considered as premature.
Knowledge of the density of states is not sufficient to discriminate between
localized and delocalized wave functions. For this we need additional
information on the two-point correlation function, the inverse participation
ratio and related quantities. The corresponding studies are under progress
and will be published separately (Guhr and St\"ockmann 2002).

If one compares the present approach with the non-linear $\sigma$ model, a dramatic
simplification is found. In the non-linear  $\sigma$ model one ends up with a
supersymmetric integral over supersymmetric field variables which can be solved only
within the zero-mode approximation. In our approach the very simple integral (\ref{46})
is obtained instead, containing only one set of supersymmetric variables, which for the
density of states even can be solved exactly. The same is true for {\em all} $n$-point
correlation functions as will be shown in Guhr and St\"ockmann (2002).

Even better, the assumption of a random distribution of point-like
scatterers applied in this work is a much more realistic description of the
situation found in mesoscopic systems than the assumption of a
delta-correlated disorder potential assumed in the non-linear $\sigma$
model.

It might be considered as a draw-back that the present derivation is based
on two unproven conjectures, namely that (i) the wave function amplitudes in
a chaotic billiard are Gaussian distributed, and that (ii) the eigenvalues
in a chaotic billiard obey random matrix behaviour. On the other hand, there
is such an overwhelming numerical evidence that both conjectures are true
that one could equally well argue that both assumptions are even better
founded than the assumption of Gaussian distributed matrix elements applied
in random matrix theory.

From the point of view of an experimentalist it would be highly desirable,
if all quantities of interest are available for systems with time-reversal
symmetry as well. Though the calculations for this case are notoriously
difficult, it should be worthwhile to undertake the effort. Experiments with
microwaves on localization-delocalization transitions, pulse propagation
etc. in disordered systems do already exist, as was mentioned in the
introduction, and wait for their proper theoretical explanation.

\ack

Intense discussions with H. Weidenm\"uller, Heidelberg, and T. Guhr, Lund, are gratefully
acknowledged. One of the referees is thanked for drawing my attention to the Lifshitz
model. This work was motivated by microwave studies of disordered systems performed by U.
Kuhl during his thesis in the author's lab. The experiments have been supported by the
Deutsche Forschungsgemeinschaft via several grants.

\appendix

\section{The density of states in the large $L$ limit}

To derive equation (\ref{72}) for the density of states we start with
equation (\ref{65}),

\begin{equation}\label{a1}
  \hat{\rho}'(z)=\frac{N!}{(L-1)!}\,e^{-z}z^{L-N-1}\,L_N^{(L-N-1)}(z)\,
  L_{N-1}^{(L-N+1)}(z)\,.
\end{equation}
where we have introduced $z=E/\lambda$ as a new variable, and where
$\hat{\rho}(z)\,dz= \rho(E)\,dE$. $z$ is assumed to be positive in the
following. Using elementary relations for the Laguerre polynomials, equation
(\ref{a1}) may be transformed as follows:

\begin{eqnarray}\label{a2}
  \hat{\rho}'(z)&=& -\frac{N!}{L!}\,e^{-z}
  \left[z^{L-N}\,L_N^{(L-N)}\right]'
  \left[L_N^{(L-N)}\right]'\nonumber\\
  &=& -\frac{N!}{L!}\,e^{-z}
  \left[e^{\frac{z}{2}}\,z^{\frac{L-N-1}{2}}y_N^{(L-N)}\right]'
  \left[e^{\frac{z}{2}}\,z^{-\frac{L-N+1}{2}}y_N^{(L-N)}\right]'\,,
\end{eqnarray}
where $y_N^{(L-N)}(z)$ is given by equation (\ref{66}). It remains to
determine the normalization constant $y_0$. From the orthogonality relation
for the Laguerre polynomials we have

\begin{equation}\label{a3}
  \int_0^\infty\left[y_n^{(\alpha)}(z)\right]^2\,\frac{\rmd z}{z}=
  \int_0^\infty e^{-z}x^\alpha\left[L_n^{(\alpha)}(z)\right]^2\,\rmd z=
  \frac{(n+\alpha)!}{n!}\,.
\end{equation}

From the WKB approximation (\ref{69}), on the other hand, we obtain

\begin{equation}\label{a4}
  \int_0^\infty\left[y_n^{(\alpha)}(z)\right]^2\,\frac{\rmd z}{z}=
  \frac{y_0^2}{2}\int_{z_0}^{z_1}\frac{\rmd z}{zq(z)}=\pi y_0^2\,,
\end{equation}
whence follows

\begin{equation}\label{a5}
  y_0=\left[\frac{1}{\pi}\frac{(n+\alpha)!}{n!}\right]^\frac{1}{2} \,.
\end{equation}

The sign has to be chosen positive to be in accordance with the usual
definition of the Laguerre polynomials. Inserting now the WKB approximation
for $y_n^{(\alpha)}(z)$ into equation (\ref{a2}) we have

\begin{equation}\label{a6}
 \hat{\rho}'(z)= -\frac{1}{\pi}\,e^{-z}
  \left[e^{\frac{z}{2}}\,z^{\frac{L-N-1}{2}}\frac{1}{\sqrt{q}}\cos w\right]'
  \left[e^{\frac{z}{2}}\,z^{-\frac{L-N+1}{2}}\frac{1}{\sqrt{q}}\cos
  w\right]'\,,
\end{equation}
where

\begin{equation}\label{a7}
  q(z)=\sqrt{\frac{L+N+1}{2z}-\frac{1}{4}-\frac{(L-N)^2}{4z^2}}\,,
\end{equation}
and

\begin{equation}\label{a8}
  w=\int_{z_0}^{z_1} q(z)\,\rmd z-\frac{\pi}{4}\,.
\end{equation}

It follows

\begin{eqnarray}\label{a9}
   \hat{\rho}'(z)&=& -\frac{1}{\pi z}
  \left[\left(\frac{1}{2}+\frac{L-N-1}{2z}-\frac{1}{2}\frac{q'}{q}\right)\frac{1}{\sqrt{q}}\cos w
  -\sqrt{q}\sin w\right]\nonumber\\
 &&\times\left[\left(\frac{1}{2}-\frac{L-N+1}{2z}
 -\frac{1}{2}\frac{q'}{q}\right)\frac{1}{\sqrt{q}}\cos w
  -\sqrt{q}\sin w\right]\,.
\end{eqnarray}

The $q'$ terms may be discarded since they are  by an order of $1/L$ smaller
as the other ones. For the same reason we may replace $L-N-1$ and $L-N+1$ by
$L-N$. Averaging over the rapidly oscillating terms, we have

\begin{equation}\label{a10}
  \hat{\rho}'(z)= -\frac{1}{2\pi z q}\left[\left(\frac{1}{2}+\frac{L-N}{2z}\right)
  \left(\frac{1}{2}-\frac{L-N}{2z}\right)+q^2\right]
\end{equation}

Inserting expression (\ref{a7}) for $q$, we obtain

\begin{eqnarray}\label{a11}
  \hat{\rho}'(z)&=&-\frac{1}{2\pi z q}\left[
  \frac{L+N}{2z}-\frac{(L-N)^2}{2z^2}\right]\nonumber\\
   &=&\frac{1}{2\pi  q}\left(q^2\right)'=\frac{1}{\pi}\,q'\,,
\end{eqnarray}

where again terms of the order of $1/L$ have been neglected. It follows
\begin{equation}\label{a12}
  \hat{\rho}(z)=\frac{1}{\pi}\,q\,,
\end{equation}

which is equivalent with equation (\ref{72}). q.\,e.\,d.

\section*{References}


\begin{harvard}

\item[] Albeverio S and \v{S}eba P 1991 {\it J. Stat. Phys.} {\bf 64} 369
\item[] Anderson P W 1978 {\it Rev. Mod. Phys.} {\bf 50} 191
\item[]  Berry M 1977 {\it J. Phys. A} {\bf 10}  2083
\item[] Bohigas O, Giannoni M and Schmit C 1984 {\it Phys. Rev. Lett.} {\bf 52}  1
\item[] Bogomolmy E, Gerland U and Schmit C 2001 {\it Phys. Rev.} {\bf E63} 36206
\item[] Efetov K 1983 {\it Adv. Phys.} {\bf 32} 53
\item[] Fal'ko V and Efetov K 1996 {\it Phys. Rev. Lett.} {\bf 77}  912
\item[] Guhr T, M\"uller-Groeling A and Weidenm\"uller H 1998 {\it Phys. Rep.
{\bf 299}}  189
\item[] Guhr T and St\"ockmann H.-J. 2002 {\it to be published}
\item[] Gutzwiller M 1990 {\it Chaos in Classical and Quantum Mechanics}, {\em
  Interdisciplinary Applied Mathematics, Vol. 1} (New York: Springer-Verlag)
\item[] Kudrolli A, Kidambi V and Sridhar S 1995 {\it Phys. Rev. Lett.} {\bf 75}  822
\item[] Lee P A and Ramakrishnan T V 1985 {\it Rev. Mod. Phys.} {\bf 57} 287
\item[] Lifshitz E M 1964 {\it Usp. Fis. Nauk.} {\bf 83} 617
\item[] \dash 1965 {\it Sov. Phys. Usp.} {\bf 7} 549 (engl. trans.)
\item[] Luttinger J M and Tao R 1983 {\it Ann. of Phys.} {\bf 145} 185
\item[] Luttinger J M and Waxler R 1987 {\it Ann. of Phys.} {\bf 175} 319
\item[] Magnus W, Oberhettinger F and  Soni R 1966 {\it Formulas and Theorems for the
  Special Functions of Mathematical Physics} (New York: Springer-Verlag)
\item[] Mehta M 1991 {\em Random Matrices. 2nd edition} (San Diego: Academic Press)
\item[] Mirlin A 2000 {\it Phys. Rep.} {\bf 326}  259
\item[] Morse P and Feshbach H 1953 {\it Methods of Theoretical Physics}
(New York: McGraw-Hill)
\item[] Soukoulis C M 1996 {\it Photonic Band Gap Materials} (Dordrecht:
Kluwer)
\item[]  Stein J, St\"ockmann H J and  Stoffregen U 1995 {\it Phys. Rev. Lett.}
{\bf 75} 53
\item[] St\"ockmann H J 1999 {\it Quantum Chaos -- An Introduction} (Cambridge:
Cambridge University Press)
\item[] St\"ockmann H J, Barth M, D\"orr U, Kuhl U and Schanze H 2001 {\it Physica E {\bf 9}}  571
\item Topinka M A, LeRoy B J, Westervelt R M, Shaw S E J, Fleischmann R, Heller E J,
Maranowski K D and Gossard A C 2001 {\it Nature} {\bf 410} 183
\item[]  Verbaarschot J, Weidenm\"uller H and Zirnbauer M 1985
{\it Phys. Rep.} {\bf 129}  367
\item[] Wiersma D S, Bartolini P, Lagendijk A and Ringhini R 1997, {\it
Nature} {\bf 390} 671
\item[] Chabanov A A and Genack A Z 2001 {\it Pys. Rev. Lett.} {\bf 87} 153901

\end{harvard}

\end{document}